\documentclass{elsart1p}

\usepackage{graphicx}
\usepackage{amssymb}

\begin{document}

\begin{frontmatter}

% Title, authors and addresses

% use the thanksref command within \title, \author or \address for footnotes;
% use the corauthref command within \author for corresponding author footnotes;
% use the ead command for the email address,
% and the form \ead[url] for the home page:
% \title{Title\thanksref{label1}}
% \thanks[label1]{}
% \author{Name\corauthref{cor1}\thanksref{label2}}
% \ead{email address}
% \ead[url]{home page}
% \thanks[label2]{}
% \corauth[cor1]{}
% \address{Address\thanksref{label3}}
% \thanks[label3]{}

\title{Non-perturbative construction of counterterms for 2PI-approximation}

% use optional labels to link authors explicitly to addresses:
% \author[label1,label2]{}
% \address[label1]{}
% \address[label2]{}

\author[1,2]{A. Patk{\'o}s}

\address[1]{Institute of Physics, E{\"o}tv{\"o}s University}
\address[2]{MTA-ELTE Research Group for Statistical and Biological
  Physics, H-1117 Budapest, Hungary}

\author[3]{Zs. Sz{\'e}p}

\address[3]{Research Institute for Solid State Physics and Optics of 
the Hungarian Academy of Sciences,\\ H-1525 Budapest, Hungary}

\begin{abstract}
% Text of abstract
A concise method is presented for the non-perturbative computation of
the counterterms renormalising 2PI-actions. The procedure is presented for
a real scalar field up to ${\cal O}(\lambda^2)$ order in the skeleton
truncation of $\Gamma_{\textrm{2PI}}$ with respect to the self-coupling, and in
a constant symmetry breaking background. The method is easily
generalizable to field theories with arbitrary global symmetry. 
\end{abstract}

\end{frontmatter}

% main text

The aim of this contribution is to provide practical tools for the
renormalisation program of the 2PI-approximate treatment of quantum
field theories. While our approach reproduces the results of previous
investigations \cite{vhees02a,blaizot03,berges05a} it allows in a
different scheme explicit determination of the counterterms. In a
broad sense the proposed method is close to the scheme of minimal
subtraction. Its detailed discussion appears in
Refs.~\cite{fejos08,patkos08}. Related results were presented by
Refs.~\cite{borsanyi08,jakovac08} and also at the conference SEWM
2008.

\section{Generic 2PI-equations}

The 2PI equations for the propagator $G(p)$ and the constant 
field expectation $v$ of a
self-interacting real scalar field adequately represent the generic
structures appearing in theories of arbitrary global symmetry:
\begin{eqnarray}
2\frac{\delta\Gamma_{\textrm 2PI}}{\delta G(p)}&=&
i G^{-1}(p)-\left[(1+\delta Z) p^2 -m^2-\delta m_2^2 
-\frac{\lambda+\delta\lambda_2}{2}v^2
- 2\frac{\delta\Gamma_2[G,v]}{\delta G(p)}\right]=0,
\nonumber\\
\frac{\delta \Gamma_{\textrm 2PI}}{\delta v}
&=&v\left[m+\delta m_0^2+\frac{\lambda+\delta\lambda_4}{6}v^2
+\frac{\lambda+\delta\lambda_2}{2}\int_p G(p) +
2 \frac{\delta\Gamma_2[G,v]}{\delta v^2}
\right]=0.
\end{eqnarray}
At finite truncation of the set of two-particle irreducible skeleton
diagrams $\Gamma_2[G,v]$ different counterterms are allowed
for each operator compatible with the symmetry of the model \cite{berges05a}.
At ${\cal O}(\lambda^2)$ \hbox{skeleton} truncation one has:
$\delta m_0^2,\delta m_2^2,\delta\lambda_0,\delta\lambda_2,$ 
$\delta\lambda_4,$ and $\delta Z.$

%\subsection
{\bf Method of renormalisation.}
In both equations one first identifies
divergent coefficients of $v^0,$ $v^2$ and 
of the finite environment dependent function 
$T_F[G]=\int_q G(q)\big|_\textnormal{finite}$.
Next, by requiring separate cancellation of these divergent coefficients
one finds the counterterms. In particular, the vanishing of the
divergent pieces  $\sim v^0$ determine $\delta m^2_i,$
that of $v^2$ determine the counterterms not related to $\Gamma_2[G,v]$
(\textit{i.e.} $\delta\lambda_2,\delta\lambda_4$). Finally,
the conditions imposed on coefficients of $T_F[G]$ 
determine counterterms defined in
$\Gamma_2[G,v]$ (\textit{i.e.} $\delta\lambda_0$, see 
Eq.~(\ref{Eq:G2_2l})).

Here we note that the method does not include discussion of overall and
sub-divergences which is an essential part of the iterative
renormalisation method.
On the other hand, the consistency of this renormalisation method 
must be checked explicitly, either analytically or numerically,
 since the number of divergence
cancellation conditions is larger than the number of counterterms.

%\subsection
{\bf Structure of the renormalised propagator.}
The expression of the self-consistent propagator contains the
self-energy split into two pieces:
\begin{equation}
i G^{-1}(p)=p^2-M^2-\Pi(p),\qquad \qquad\Pi(p)=\Pi_a(p)+\Pi_0(p)+\Pi_r(p),
\end{equation}
where 
$M^2=m^2+\frac{\lambda}{2} v^2+\frac{\lambda}{2} T_F[G]$ 
is the momentum independent local part, and
the momentum dependent (non-local) part $\Pi(p)$ itself is further 
decomposed into pieces displaying different asymptotics for large $p$:
$\Pi_a(p)\sim p^2 (\ln p)^{c_1}, \Pi_0(p)\sim (\ln p)^{c_2},
\Pi_r(p)\sim p^{-2}$.

The divergences are separated with help of an {\it auxiliary propagator} 
$G_a(p)$ which has the same asymptotic behaviour as $G(p)$ and 
invokes also an arbitrarily chosen IR regulator $M_0^2$ in its definition: 
\begin{equation}
i G_a^{-1}(p)=p^2-M_0^2-\Pi_a(p).
\end{equation}

\section{2-loop truncation of the effective action: 
$\Gamma_2^{(2l)}[G,v]$}

Renormalisation of this truncation is simpler and still might have
several physically interesting applications 
\cite{andersen08,tranberg08}.  
The contribution of the following  2PI diagrams \cite{arrizabalaga06}
\begin{equation}
\Gamma_2^{(2l)}=\frac{\lambda+\delta\lambda_0}{8} 
\raisebox{-0.38cm}{\includegraphics[width=0.5cm]{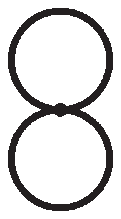}} 
+ \frac{\lambda^2}{12} v^2 
\raisebox{-0.38cm}{\includegraphics[width=0.95cm]{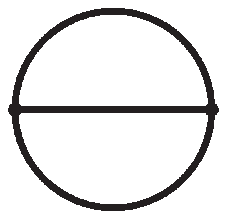}}
\label{Eq:G2_2l}
\end{equation}
to the propagator and field equations is
\begin{equation}
2\frac{\delta \Gamma_2^{(2l)}[G,v]}{\delta G(p)}=
\frac{1}{2}(\lambda+\delta\lambda_0)T[G]+
\frac{1}{2}\lambda^2 v^2 I(p,G),\qquad
2\frac{\delta \Gamma_2^{(2l)}[G,v]}{\delta v^2}=\frac{\lambda^2}{6}S(0,G),
\end{equation}
where we define 
$I(p,G)=-i\int_q G(q) G(p+q),  S(0,G)=\int_p G(p) I(p,G)$.

In the present case the form of the auxiliary propagator is simple,
since the leading asymptotics of the self-consistent propagator is unchanged:
$i G_a^{-1}=p^2-M_0^2$, ($\Pi_a\equiv 0$).
The separation of divergences is achieved by expanding the self-consistent
propagator $G(p)$ around $G_a(p)$: 
\begin{eqnarray}
I_\textnormal{div}^{(2l)}(p,G)&=&T_a^{(0)}:=
\raisebox{-0.16cm}{\includegraphics[width=0.7cm]{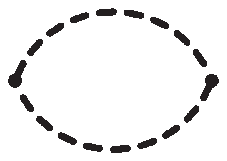}}
\big|_\textnormal{div},\quad
T_\textnormal{div}^{(2l)}[G] =T_a^{(2)} + 
(M^2-M_0^2) T_a^{(0)}+ \frac{\lambda^2}{2}v^2 T_a^{(I)},\\
S_\textnormal{div}^{(2l)}(0,G)&=&\raisebox{-0.28cm}
{\includegraphics[width=0.8cm]{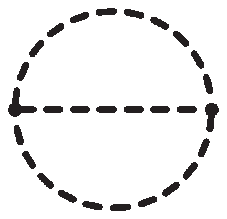}}\Big|_\textnormal{div}
+\frac{3}{2}\lambda^2 v^2\left[ T_a^{(0)} T_a^{(I)} + T_a^{(I,2)}
\right]\nonumber\\
&&
\quad\quad\quad\quad\,\,+T_a^{(0)}T_F+
3(M^2-M_0^2) \left[\big(T_a^{(0)}\big)^2 + T_a^{(I)} \right]. 
\end{eqnarray}
The lines in the diagrams refer to $G_a$ with the following graphical 
notations (the ``bubble'' with letter `F' inside denotes the finite part of 
the auxiliary bubble integral):
\begin{equation}
T_a^{(2)}=\raisebox{0.0cm}{\includegraphics[width=0.70cm]{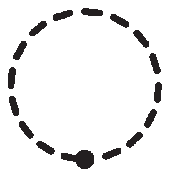}} 
\raisebox{0.22cm}{$\Big|_\textnormal{div}$},\quad 
T_a^{(I)}=\raisebox{0.0cm}{\includegraphics[width=0.70cm]{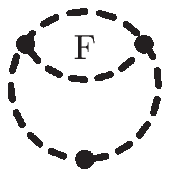}} 
\raisebox{0.22cm}{$\Big|_\textnormal{div}$},\quad  
T_a^{(I,2)}=\raisebox{0.0cm}{\includegraphics[width=0.70cm]{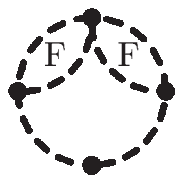}} 
\raisebox{0.22cm}{$\Big|_\textnormal{div}$}.
\nonumber
\end{equation}

Since $M^2$ includes $T_F[G],$ its presence leads to the danger of 
{\it environment dependent divergences} in 
$T_\textnormal{div}^{(2l)}[G]$ and $S_\textnormal{div}^{(2l)}(0,G).$
The conditions for the cancellation of such divergences in the propagator and 
the field equation are, respectively:
\begin{equation}
%&&
\delta\lambda_0+\frac{\lambda}{2}(\lambda+\delta\lambda_0)
T_a^{(0)}=0,\quad
%\nonumber\\
%&&
\delta\lambda_2+\frac{
\lambda}{2}(3\lambda+\delta\lambda_2) T_a^{(0)}
+\frac{\lambda^3}{2}\left[\big(T_a^{(0)}\big)^2 + T_a^{(I)} \right]=0.
\end{equation}
On the other hand, the $v^2$-dependent divergence cancellation in the propagator
gives a relation between 
$\delta\lambda_0$ and $\delta\lambda_2$:
\begin{equation}
\delta\lambda_2+\frac{1}{2}\lambda(\lambda+\delta\lambda_0)
\left(T_a^{(0)}+\lambda T_a^{(I)}\right)=0.
\end{equation}
By the previously determined $\delta\lambda_0$ and $\delta\lambda_2$ 
this  \textit{consistency relation} is satisfied.

In turn the cancellation of the further 
two $v^0$-dependent plus one $v^2$-dependent divergences 
determines $\delta m_0^2,\delta m_2^2,$ and $\delta\lambda_4.$

\section{Adding the basket-ball to the effective action}

Now one completes the 2PI-part of the action with the ``basket-ball''
diagram:
 \begin{equation}
\Gamma_2^{\lambda\!^2}[G,v]=\Gamma_2^{(2l)}[G,v]
+ \frac{\lambda^2}{48} 
\raisebox{-0.31cm}{\includegraphics[width=0.85cm]{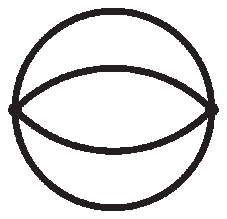}}.
\end{equation}
The form of the equation of state remains unchanged, while the
self-energy receives the contribution:
$\Pi_2(p)=\frac{\lambda^2}{6}S_F(p,G)$, which results in 
$\Pi_a(p)\equiv\Pi_{2,a}(p)\neq 0$ term.

The analysis reveals, for instance, the following
form for the divergent part of the tadpole integral:
\begin{equation}
T_\textnormal{div}[G]=T_\textnormal{div}^{(2l)}[G]-i\int_k
G_a^2(k)\Pi_{2,0}(k).
\end{equation}
Similar (but more complicated) expression is derived for $S_\textrm{div}(p,G)$.
Therefore, the crucial object to be found for the explicit
counterterm construction is the subleading asymptotic piece:
$\Pi_{2,0}$, which is the logarithmic part of $\Pi_2(p)$.
An integral representation can be derived for it:
\begin{equation}
\label{inteq}
\Pi_{2,0}(p)=-\frac{i}{2} \int_k G_a^2(k) K(p,k)\left[M^2-M_0^2
+ \frac{\lambda^2}{2} v^2I_a^F(k)+\Pi_{2,0}(p)\right],
\end{equation}
with the kernel defined with help of the renormalised bubble-integrals
\cite{patkos08}:
\begin{equation}
K(p,k)=\frac{\lambda^2}{2}\left[I_a^F(k+p)+I_a^F(k-p)-2I_a^F(k)\right].
\end{equation}

Solving (\ref{inteq}) for $\Pi_{2,0}(p)$ one finds a linear
combination of $M^2-M_0^2$ and $\lambda^2 v^2$:
\begin{equation}
\Pi_{2,0}(p)=\frac{1}{2}(M^2-M_0^2)\Gamma_0(p)+
\frac{1}{4} \lambda^2 v^2 \Gamma_1(p), 
\end{equation}
where
$
\Gamma_0(p) =-i \int_k\Gamma(p,k)G_a^2(k),
\Gamma_1(p)= -i\int_k \Gamma(p,k)G_a^2(k)I_a^F(k).$
The kernel which determines the coefficient functions $\Gamma_0(p),
\Gamma_1(p)$ satisfies a finite Bethe--Salpeter-type equation: 
\begin{equation}
\Gamma(p,k)=K(p,k)-\frac{i}{2}\int_q G_a^2(q)K(p,q)\Gamma(q,k).
\end{equation}

This result demonstrates that by adding the basket-ball the types of
the occurring divergent coefficients do not change 
(e.g. $\sim v^0, v^2, T_F$). The same procedure, as described for 
the 2-loop truncation allows also the construction for the counterterms 
up to ${\cal O}(\lambda^2)$ accuracy. The verification of the consistency 
of the counterterm determination with the redundant conditions becomes 
more cumbersome. Applicability of this renormalisation procedure together 
with the analytical check of some of the consistency relations to 
multicomponent scalar models was illustrated on the $O(N)$ model in 
Ref.~\cite{patkos08}.

\section*{Acknowledgements}
Work supported by the Hungarian Scientific Research Fund
under contracts T-046129 and
T-068108. Zs. Sz. is supported by OTKA Postdoctoral Grant
No. PD050015.
 
% The Appendices part is started with the command \appendix;
% appendix sections are then done as normal sections
% \appendix

% \section{}
% \label{}

\end{document}